\documentclass[twoside]{ilcws07}
\usepackage[latin1]{inputenc}
\usepackage[dvips]{graphicx,epsfig,color}
\usepackage{wrapfig,rotating}
\usepackage{amssymb,amsmath,array}

\begin{document}
\title{CP Violation in SUSY Particle Production and Decay} %% 
%***********************************************************************
% AUTHORS INFORMATION AREA
%***********************************************************************
\author{Stefan Hesselbach
% Optional short acknowledgment: remove next line if non-needed
%\thanks{This is an optional funding source acknowledgment.}
% DO NOT MODIFY THE FOLLOWING '\vspace' ARGUMENT
\vspace{.3cm}\\
% Addresses and institutions (remove "1- " in case of a single institution)
University of Southampton - School of Physics \& Astronomy \\
Highfield, Southampton SO17 1BJ - UK
%% Remove the next three lines in case of a single institution
}
%%***********************************************************************
% END OF AUTHORS INFORMATION AREA
%***********************************************************************

\maketitle

\begin{abstract}
Recent studies about CP violation in the 
Minimal Supersymmetric Standard Model (MSSM)
with complex parameters are reviewed.
In order to unambiguously identify the CP-violating phenomena it is
necessary to study CP-odd or T-odd observables.
In chargino and neutralino production and decay at the 
International Linear Collider (ILC)
triple product asymmetries and asymmetries defined via
transverse beam polarization have been analyzed.
It has been found that these asymmetries
can be measured at the ILC in a large region of the MSSM
parameter space and are thus an important tool to establish CP violation
in supersymmetry.
\end{abstract}

\section{Introduction}

In the Lagrangian of the Minimal Supersymmetric Standard Model (MSSM) many 
parameters can be complex which can give rise to new CP-violating
phenomena~\cite{Ibrahim:2007fb}
and may help to explain the baryon-antibaryon asymmetry of
the universe~\cite{BAU}.
After the elimination of unphysical phases two complex parameters remain in
the neutralino and chargino sector, the U(1) gaugino mass parameter
$M_1$ and the higgsino mass parameter $\mu$, whereas the SU(2) gaugino
mass parameter $M_2$ and the ratio $\tan\beta$ of the Higgs vacuum expectation
values can be chosen real and positive.
In addition the SU(3) gaugino (gluino) mass parameter $M_3$
and the trilinear scalar couplings $A_f$ in the sfermion sector 
can be complex.

The new CP-violating phases are constrained by the experimental bounds
on electric dipole moments (EDMs) of electron, neutron and Hg atom.
However, these constraints are highly model-dependent.
In constrained MSSM scenarios only small 
values of the phases are allowed, especially the phase of $\mu$
is strongly limited.
In more general supersymmetric (SUSY) models larger phases may be possible
due to cancellations between different SUSY contributions to
the EDMs
or in SUSY models with heavy sfermions in the first two
generations~\cite{EDM}.
For instance, it has been pointed out recently that for large $A_f$,
phases $\phi_\mu\sim O(1)$ can be compatible with
the EDM constraints~\cite{Yaser Ayazi:2006zw}.
Furthermore, the restrictions on the phases may also disappear if
lepton flavor violating terms in the MSSM Lagrangian are
included~\cite{Bartl:2003ju,Ayazi:2007kd}.
In conclusion, large phases of SUSY parameters cannot be ruled out by
present EDM experiments.

The precise determination of the underlying SUSY parameters 
including the phases is an important task of the International Linear
Collider (ILC)~\cite{Brau:2007zz}.
The parameters $M_1$, $M_2$, $\mu$ and $\tan\beta$ of the neutralino
and chargino sector are expected to be determined with very high
precision which can be further enhanced by combining LHC and ILC
analyses~\cite{Weiglein:2004hn}.
The impact of the SUSY CP phases on the MSSM Higgs sector is
summarized in~\cite{Accomando:2006ga}.
While CP-even observables like production cross sections and decay
branching ratios may strongly depend on the new phases,
CP-odd observables are necessary to unambiguously determine the phases
and establish CP violation~\cite{Hesselbach:2004sp}.
Concerning CP-even observables especially the decays of SUSY particles and
Higgs bosons are a sensitive probe of the SUSY phases~\cite{DECAY}.
CP-odd observables can be constructed in form of rate asymmetries or
with the help of triple products, transverse beam polarization or the
polarization of final state particles, for recent studies see
e.g.~\cite{CPodd}.
In this contribution studies about CP-odd triple product asymmetries
and asymmetries defined via transverse beam polarization in chargino
and neutralino production and decay at the ILC are reviewed, focusing
especially on their measurability.

\section{Triple product asymmetries}

T-odd triple product correlations between momenta and spins of the
involved particles allow the definition of CP-odd asymmetries already
at tree level~\cite{tripleproducts}.
For chargino and neutralino production and subsequent two-body decays
CP-odd and T-odd asymmetries based on triple products and their
measurability have been thoroughly studied in~\cite{AT2body}.
Decays involving $W$ and $Z$ bosons and those into sfermions and
fermions have been analyzed and it has been found that especially in
the latter case large asymmetries up to $30\%$ are possible.

Here, I will focus on two studies about chargino and neutralino production
and subsequent three-body decays~\cite{Bartl:2006yv,Bartl:2004jj},
$e^+ e^- \to \tilde{\chi}_i + \tilde{\chi}_j \to
\tilde{\chi}_i + \tilde{\chi}^0_1 f \bar{f}^{(')}$.
Including full spin correlations between production and decay
products of the form
$i\epsilon_{\mu\nu\rho\sigma}p^\mu_ip^\nu_jp^\rho_kp^\sigma_l$
(where the $p^\mu_i$ denote the momenta of the involved particles)
appear in the amplitude squared
in terms, which depend on the spin of the decaying chargino
or neutralino~\cite{spincorr}.
Together with the complex parameters entering the
couplings these terms can give
real contributions to suitable observables at tree-level.
Triple products
$\mathcal{T}_1 = \vec{p}_{e^-}\cdot(\vec{p}_{f}\times\vec{p}_{\bar{f}^{(')}})$
of the initial electron momentum $\vec{p}_{e^-}$ and
the two final fermion momenta $\vec{p}_{f}$ and $\vec{p}_{\bar{f}^{(')}}$
or
$\mathcal{T}_2 = \vec{p}_{e^-}\cdot(\vec{p}_{\tilde{\chi}_j}\times\vec{p}_{f})$
of the initial electron momentum $\vec{p}_{e^-}$, the momentum of the
decaying neutralino or chargino $\vec{p}_{\tilde{\chi}_j}$ and one
final fermion momentum $\vec{p}_{f}$
allow the definition of T-odd asymmetries
\begin{equation}
A_T = \frac{\sigma({\cal T}_i>0) - \sigma({\cal T}_i<0)}%
 {\sigma({\cal T}_i>0) + \sigma({\cal T}_i<0)}
 =
 \frac{\int {\rm sign}({\cal T}_i) |T|^2 d{\rm Lips}}%
 {{\int}|T|^2 d{\rm Lips}},
\end{equation}
where ${\int}|T|^2 d{\rm Lips}$ is
proportional to the cross section $\sigma$ of the 
combined production and decay process.
$A_T$ is odd under naive time-reversal operation and hence CP-odd, if
higher order
final-state interactions and finite-widths effects can be neglected.
In the case of chargino production and decay where the asymmetry
$\bar{A}_T$ for the charge-conjugated process is accessible
a genuine CP asymmetry
\begin{equation}\label{ACP}
A_{\rm CP} = \frac{A_T - \bar{A}_T}{2}
\end{equation}
can be defined.

The statistical significance $S$ to which above asymmetries can be
determined to be non-zero can be estimated in the following way:
The absolute error of $A_T$ is given by
$\Delta A_T = S \sqrt{1 - A_T^2}/\sqrt{\sigma \mathcal{L}_\mathrm{int}}$,
where $S$ denotes the number of standard
deviations, $\sigma$ the cross section of the respective process
and $\mathcal{L}_\mathrm{int}$ the integrated luminosity~\cite{Desch:2006xp}.
For $A_T \lesssim 10\%$ it is 
$\Delta A_T = S/\sqrt{\sigma \mathcal{L}_\mathrm{int}}$
in good approximation and requiring $A_T > \Delta A_T$ for $A_T$ to be
measurable one obtains
\begin{equation} \label{nsigma}
S = \sqrt{A_T^2 \sigma \mathcal{L}_\mathrm{int}}
\qquad \textrm{and} \qquad
S = \sqrt{2 A_{\rm CP}^2 \sigma \mathcal{L}_\mathrm{int}}\,,
\end{equation}
respectively,
assuming that the statistical errors of $A_T$ and $\bar{A}_T$ are
independent of each other.
$S$ can be used as an estimation of the measurability of the asymmetries.
However, in order to determine the final accuracy in the experiment also
initial state radiation, beamstrahlung, backgrounds and detector effects
have to be included. For neutralino production and decay this has
been analyzed in~\cite{AguilarSaavedra:2004dz} and it has been found
that asymmetries $\mathcal{O}(10\,\%)$ are detectable
after few years of running of the ILC.

\begin{figure}[t!]
\centerline{\epsfig{file=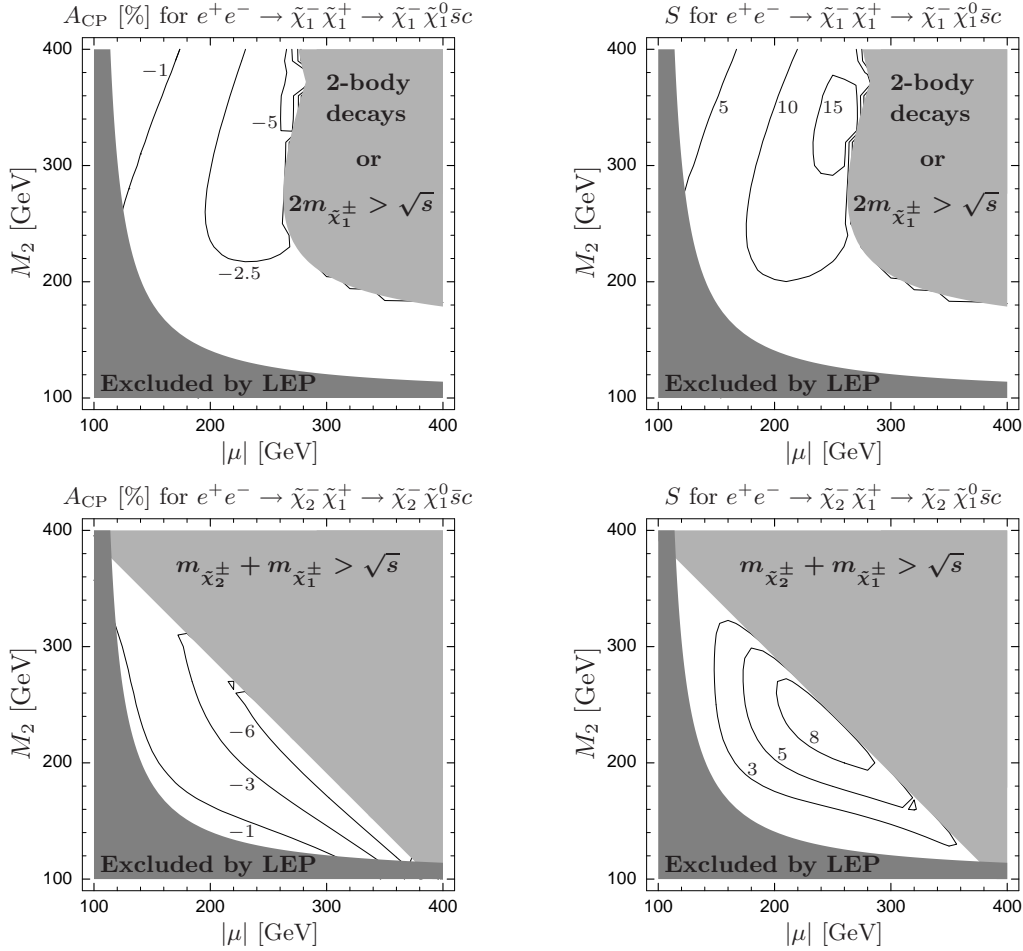}}
\caption{Contour lines of the CP-odd triple product asymmetry
$A_\mathrm{CP}$, Eq.~(\ref{ACP}),
and statistical significance $S$ using
$\mathcal{T}_1 = \vec{p}_{e^-}\cdot(\vec{p}_{\bar{s}}\times\vec{p}_{c})$
for
$|M_1|/M_2= 5/3 \tan^2\theta_W$,
$\phi_{M_1}=0.5\pi$, $\phi_{\mu}=0$,
$\tan\beta = 5$, $m_{\tilde{\nu}} = 250$~GeV,
$m_{\tilde{c}} = 500$~GeV, $m_{\tilde{s}} = 505.9$~GeV,
$\sqrt{s}=500$~GeV, $\mathcal{L}_\mathrm{int}=500~\mathrm{fb}^{-1}$
and longitudinal beam polarizations $(P_{e^-},P_{e^+}) = (-80\%, +60\%)$.
From~\cite{Bartl:2006yv}.}
\label{Fig:ACPchargino}
\end{figure}

In Figure~\ref{Fig:ACPchargino} $A_\mathrm{CP}$ and $S$
are shown for chargino production
$e^+ e^- \to \tilde{\chi}_j^- \tilde{\chi}_1^+$, $j=1,2$ and subsequent decay
$\tilde{\chi}_1^+ \to \tilde{\chi}^0_1 \bar{s} c$
using the triple product
$\mathcal{T}_1 =
\vec{p}_{e^-}\cdot(\vec{p}_{\bar{s}}\times\vec{p}_{c})$~\cite{Bartl:2006yv}.
Note that the statistical significance $S$ is larger than $5$ in large
regions of the parameter space. However, in order to measure
$A_\mathrm{CP}$ it is necessary to discriminate the two outgoing quark
jets, i.e.\ to tag the $c$ jet. The respective $c$ tagging efficiency
will decrease the final significance by about a factor $0.5$ but
nevertheless large regions of the parameter space can be covered.
If instead the production plane is reconstructed by analyzing the
decays of the  $\tilde{\chi}_2^-$ in
$e^+ e^- \to \tilde{\chi}_2^- \tilde{\chi}_1^+$
also the leptonic decays 
$\tilde{\chi}_1^+ \to \tilde{\chi}^0_1 \ell^+ \nu$
can be used to define $A_\mathrm{CP}$
using
$\mathcal{T}_2 =
\vec{p}_{e^-}\cdot(\vec{p}_{\tilde{\chi}_1^+}\times\vec{p}_{\ell^+})$.
In this case, however, the cross sections are rather small, hence $S$
is always smaller than about $5$ despite potentially large asymmetries
$A_\mathrm{CP}$.

For neutralino production
$e^+e^-\rightarrow\tilde{\chi}^0_i\tilde{\chi}^0_2$, $i=1,\ldots,4\,$,
with subsequent leptonic three-body decay
$\tilde{\chi}^0_2\rightarrow\tilde{\chi}^0_1\ell^+\ell^-$,
$\ell=e,\mu$, the triple product
$\mathcal{T}_1 = \vec{p}_{e^-}\cdot(\vec{p}_{\ell^+}\times\vec{p}_{\ell^-})$
can be used to define the T-odd asymmetry $A_T$, which is directly
measurable without reconstruction of the momentum of the decaying neutralino
or further final-state analyses. 
It has been found in~\cite{Bartl:2004jj} that $A_T \gtrsim 10\%$ in large
regions of the parameter space for 
$e^+ e^- \to \tilde{\chi}^0_j + \tilde{\chi}^0_2 \to
\tilde{\chi}^0_j + \tilde{\chi}^0_1 \ell^+ \ell^-$, $j=1,3$,
yielding significances $S$ larger than 5.

\section{Asymmetries using transverse beam polarization}

The use of transverse beam polarization offers further possibilities to
define CP-sensitive observables.
As all terms in the squared amplitude $|T|^2$ of respective processes which are
sensitive to transverse beam polarization depend on the
product of the degrees of transverse beam polarization of both beams the
CP-sensitive observables are only accessible if both beams of the ILC can be
polarized~\cite{MoortgatPick:2005cw}.
The respective terms in $|T|^2$ contain products of the form
$i\epsilon_{\mu\nu\rho\sigma} t_\pm^\mu p^\nu_i p^\rho_j p^\sigma_k$
or
$i\epsilon_{\mu\nu\rho\sigma} t_+^\mu t_-^\nu p^\rho_i p^\sigma_j$,
where $t_\pm^\mu$ is the 4-vector of the transverse beam
polarization of the positron and electron
beam\footnote{For a detailed definition see e.g.~\cite{Bartl:2005uh}.},
respectively,
and the $p^\nu_i$ denote the momenta of the involved particles.
This in turn allows the definition of CP-odd asymmetries in suitable
production and decay processes.
In~\cite{Bartl:2006bn} such asymmetries and their measurability
have been analyzed for selectron production at an $e^- e^-$ collider.
In~\cite{Choi:2006vh,Bartl:2007qy}
CP-odd asymmetries using transverse beam polarization
have been studied for neutralino production and subsequent
two-body decays and their measurability has been
compared with CP asymmetries accessible with unpolarized
or longitudinally polarized beams.

In chargino production $e^+ e^- \to \tilde{\chi}^+_i \tilde{\chi}^-_j$ 
all CP-odd terms in $|T|^2$ vanish because of CPT invariance and the
fact that charginos are Dirac particles~\cite{Bartl:2004xy}.
Due to the Majorana nature of the neutralinos the respective terms are allowed
in neutralino production $e^+ e^- \to \tilde{\chi}^0_i \tilde{\chi}^0_j$
and CP-odd asymmetries can be defined by analyzing the azimuthal distributions
of the neutralinos~\cite{Bartl:2005uh}:
\begin{equation} \label{ACPtrans}
A_\mathrm{CP}=
\left[\int^{\pi/2}_{0} -\int^{\pi}_{\pi/2}\right]
A_{CP}(\theta)\,{\rm d}\theta \, ,
\end{equation}
\begin{equation}
A_\mathrm{CP}(\theta)=\frac{1}{\sigma}
\left[
  \int^{\frac{\pi}{2}+\frac{\eta}{2}}_{\frac{\eta}{2}}-
  \int^{\pi+\frac{\eta}{2}}_{\frac{\pi}{2}+\frac{\eta}{2}}+
  \int^{\frac{3\pi}{2}+\frac{\eta}{2}}_{\pi+\frac{\eta}{2}}-
  \int^{2\pi+\frac{\eta}{2}}_{\frac{3\pi}{2}+\frac{\eta}{2}}
\right]
\frac{{\rm d}^2\sigma}{{\rm d}\phi \, {\rm d}\theta} {\rm d}\phi \, ,
\end{equation}
where $\phi$ denotes the azimuthal angle of the scattering plane and
$\eta$ the orientation of the transverse polarizations.
The statistical significance is given by
$S = \sqrt{A_\mathrm{CP}^2 \sigma \mathcal{L}_\mathrm{int}}$ or
vice versa the
necessary integrated luminosity to reach a certain significance by
$\mathcal{L}_\mathrm{int} = S^2/(A_\mathrm{CP}^2 \sigma)$.

\begin{figure}[t!]
\centerline{\epsfig{file=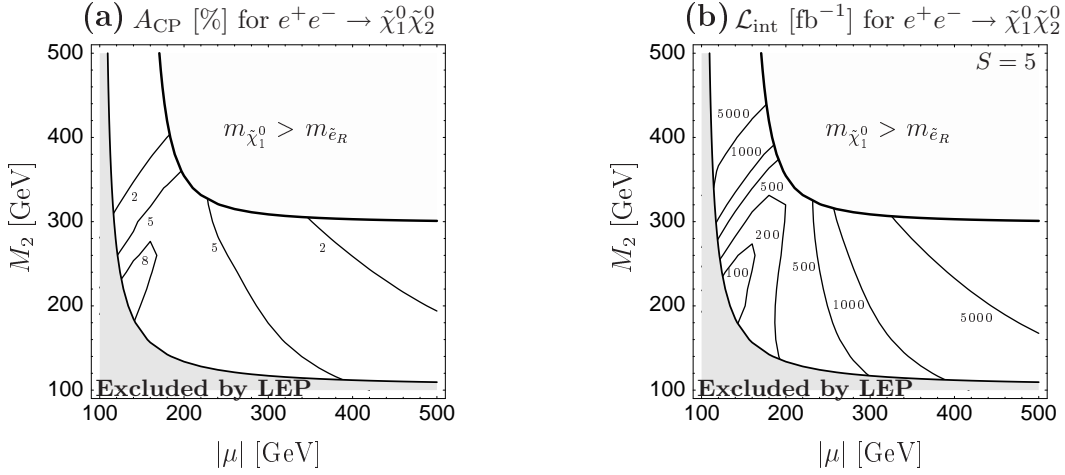}}
\caption{Contour lines of (a) the
CP-odd asymmetry $A_\mathrm{CP}$, Eq.~(\ref{ACPtrans}),
defined with help of transverse beam
polarization
and (b) the necessary integrated luminosity to
reach a significance $S=5$ for 
$|M_1|/M_2 = 5/3 \tan^2\theta_W$,
$\phi_{M_1}=0.5 \pi$, $\phi_{\mu}=0$,
$\tan\beta = 5$,
$m_{\tilde{e}_L} = 400~\mathrm{GeV}$,
$m_{\tilde{e}_R} = 150~\mathrm{GeV}$,
$\sqrt{s} = 500~\mathrm{GeV}$ and degrees of transverse beam polarizations
of (a) $(P_{e^-}^T,P_{e^+}^T) = (100\%, 100\%)$ 
and (b) $(P_{e^-}^T,P_{e^+}^T) = (80\%, 60\%)$.
From~\cite{Bartl:2005uh}.
}
\label{Fig:ACPtransverse}
\end{figure}

In Figure~\ref{Fig:ACPtransverse} $A_\mathrm{CP}$ and
$\mathcal{L}_\mathrm{int}$ necessary to reach $S=5$ are shown for 
$e^+ e^- \to \tilde{\chi}^0_1 \tilde{\chi}^0_2$, where it can be seen that
also the CP-odd asymmetry defined via transverse beam polarization can be
measured in large regions of the SUSY parameter space at the
ILC~\cite{Bartl:2005uh}.
Similarly, the respective asymmetries for
$e^+ e^- \to \tilde{\chi}^0_1 \tilde{\chi}^0_3$ are well measurable in large
regions of the parameter space.
However, in order to measure $A_\mathrm{CP}$ the production plane has to be
reconstructed. This is not necessary if the subsequent decays of the
neutralinos are included. It has been found in~\cite{Bartl:2005uh,Bartl:2007qy}
that respective asymmetries including two-body decays of the neutralinos are
also measurable in large regions of the SUSY parameter space.

\section{Conclusions}

Recent studies analyzing CP-odd or T-odd
triple product asymmetries or asymmetries
defined via transverse beam polarization in chargino and neutralino production
and decay have been reviewed.
It has been found that these asymmetries
are measurable in large regions of the SUSY parameter space and
are thus an important tool to search for CP violation in SUSY and to
unambiguously determine the SUSY phases.

% ****************************************************************************
% BIBLIOGRAPHY AREA
% ****************************************************************************

\begin{footnotesize}
% IF YOU DO NOT USE BIBTEX, USE THE FOLLOWING SAMPLE SCHEME FOR THE REFERENCES
% ----------------------------------------------------------------------------

% ----------------------------------------------------------------------------

\end{footnotesize}

% ****************************************************************************
% END OF BIBLIOGRAPHY AREA
% ****************************************************************************

\end{document}